\documentclass[aps,prx,twocolumn,notitlepage,superscriptaddress]{revtex4-2}
\usepackage{amsmath,amssymb}
\usepackage[hidelinks,colorlinks,linkcolor=blue,
citecolor=blue,urlcolor=blue]{hyperref}
\usepackage{graphicx,siunitx}
\usepackage[dvipsnames]{xcolor}

\usepackage{pifont}

\usepackage[normalem]{ulem}

\usepackage{amsmath,amsfonts,amssymb,amsthm,epsfig,array}
\usepackage{dsfont}
\usepackage{slashed}
\usepackage{graphics}
\usepackage{float}
\usepackage{verbatim}
\usepackage{color}
\usepackage{tabularx}
\usepackage[mathscr]{euscript}
\usepackage{mathtools}
\usepackage{braket}
\usepackage{physics}

\newcommand{\PRLsec}[1]{\emph{#1---}}
\newcommand{\bK}{\mathbf{K}}
\newcommand{\bk}{\mathbf{k}}
\newcommand{\bq}{\mathbf{q}}
\newcommand{\bp}{\mathbf{p}}

\newcommand{\bG}{\mathbf{G}}

\newcommand{\br}{\mathbf{r}}
\newcommand{\bA}{\mathbf{A}}
\newcommand{\ba}{\mathbf{a}}
\newcommand{\myw}{w}

\begin{document}
\title{Interacting phase diagram of twisted bilayer MoTe$_2$ in magnetic field}

\author{Minxuan Wang}
\affiliation{National High Magnetic Field Lab, Tallahassee, FL 32310}
\affiliation{Department of Physics, University of California, Berkeley, CA 94720, USA}

\author{Xiaoyu Wang}
\affiliation{National High Magnetic Field Lab, Tallahassee, FL 32310}

\author{Oskar Vafek}
\affiliation{National High Magnetic Field Lab, Tallahassee, FL 32310}
\affiliation{Department of Physics, Florida State University, Tallahassee, FL 32306}

\begin{abstract}
We study electron-electron interaction induced states of twisted bilayer MoTe$_2$ in an out-of-plane magnetic field $B\hat{\bf z}$ near one hole per moir\'e unit cell filling.
The 3D phase diagram showing the evolution of competing phases with $B$, interaction strength and an out-of-plane electric field is presented at electron fillings that follow the Diophantine equation along Chern number $-\text{sign}\left(B\right)$ line, that is pointing away from the charge neutral filling, where we find prominent Chern insulators consistent with the experiments.
We also explain the experimental absence of prominent Chern insulators along the Chern number $+\text{sign}\left(B\right)$ line.
\end{abstract}

\maketitle
\PRLsec{Introduction.} Moir\'e materials are well known for their tunability and hosting various correlated phases due to their narrow energy electronic bands\cite{balents2020superconductivity,kennes2021moire}. Among these materials, twisted transition metal dichalcogenides (TMD) homobilayers, as well as heterobilayers, are predicted\cite{wu2019topological,devakul2021magic}, and experimentally shown, to exhibit integer quantum anomalous Hall (QAH) effect \cite{li2021quantum, xie2022valley}. Intriguingly, fractional quantum anomalous Hall effect has recently also been observed in twisted MoTe$_2$ homobilayer (tMoTe$_2$) \cite{cai2023signatures, zeng2023thermodynamic, park2023observation, xu2023observation}, with a number of theoretical papers devoted to these integer and fractional Chern insulators\cite{TitusPRL2011, Sheng_2011, RegnaultBernevigPhysRevX2011, TangPRL2011, KaiSunPRL2011,  abouelkomsan2023multiferroicity, wang2023fractional, crepel2023anomalous, yu2024fractional, goldman2023zero, qiu2023interaction, wang2023topological, wang2023topology, fan2024orbital, liu2023gate, luo2024majorana, li2024electrically}.
Due to the large unit cell area $A_{uc}$ of moir\'e materials, out-of-plane magnetic field $B\hat{\bf z}$ has proven to be a powerful probe of their correlated states \cite{lu2021multiple, foutty2023mapping, cai2023signatures, zeng2023thermodynamic}. Particularly interesting is the evolution of such states in the $\nu$-$B$ plane, where the electron filling $\nu=\rho_eA_{uc}$ and $\rho_e$ is the areal density of electrons.
The key tool is the thermodynamic formula \cite{Streda1982, Widom1982} which states that the quantized Hall conductivity of a gapped state of electrons (with charge $-e$) at $T=0$ follows
\begin{equation}\label{eq:Streda}
\sigma_{xy}=-\sigma_{yx}=-e\frac{\partial\rho_e}{\partial B}\bigg|_{\mu,T}=-t\frac{e^2}{h}.
\end{equation}
Here $t$ is the Chern number\cite{Kohmoto1985} and we use the SI units throughout.
The Diophantine equation for $\nu$ is\cite{Wannier1978,Thouless1983,MacDonald1983,Tesanovic1989, spanton2018observation}
\begin{equation}\label{eq:Diophantine}
\nu=s+t\frac{\phi}{\phi_0},
\end{equation}
where the flux quantum $\phi_0=h/e=4136 \mathrm{T\cdot nm}^2$ and the flux through the unit cell is $\phi=BA_{uc}$.

A ubiquitous experimental observation in tMoTe$_2$ at twist angles in the $3.4^\circ \sim 3.9^\circ$ range
is the QAH state at $\nu=-1=s$ evolving into a Chern state at $B\neq0$ with a prominent gap at $t=-\text{sign}(B)$ and a noted absence of a prominent gap at $t=+\text{sign}(B)$ \cite{cai2023signatures,zeng2023thermodynamic,park2023observation, xu2023observation}. In other words, the Streda lines emanating from $\nu=-1$ point {\em away} from the charge neutrality point (CNP) for either sign of $B$. The two QAH states with opposite Chern numbers at $B=0$ are partners under the time reversal symmetry, which is spontaneously broken, and therefore they must be equally stable at $B=0$.
The relative instability of the $t=+\text{sign}(B)$ state at $B\neq0$ has thus far not been explained. Note that in $1.23^\circ$ twisted WSe$_2$ bilayer the prominent Chern insulator line at $s=-1$ points {\em towards} the CNP\cite{foutty2023mapping}, and that {\em both} behaviors have been observed in the magic angle twisted bilayer graphene device of Ref.\cite{Xie2021}, as well as in calculations \cite{wang2023theory}.

Here we show that, within our $B\neq0$ self-consistent Hartree-Fock calculation at the twist angle $3.89^\circ$ and the dielectric constant $\epsilon$ entering the interaction (\ref{eq:myInteraction}) chosen to reproduce the $B=0$ QAH state\cite{wang2023topology}, the ground state at $s=-1$ and $t=-\text{sign}(B)$ is valley polarized and that the holes preferentially populate the valley whose spin is aligned to the direction of the magnetic field. This state can be thought of as a Landau quantized QAH state with a non-vanishing (large) gap in the $B\rightarrow 0$ limit and we refer to it as a Chern paraelectric because it preserves $C_{2y}\mathcal{T}$, the combination of an in-plane 2-fold rotation symmetry and the time reversal symmetry. Its energy spectrum as a function of $\phi/\phi_0$ is shown in the Fig.\ref{fig:my2}(a). Since the spin orientation of the holes in the two valleys is opposite due to the
spin-valley locking\cite{DiXiaoPRL2012}, the valley polarization reverses upon reversing the sign of $B$ (still at $t=-\text{sign}(B)$). Interestingly, the nature of the lowest energy state at $s=-1$ and $t=+\text{sign}(B)$ depends sensitively on $g$, the strength of the spin and orbital Zeeman coupling\cite{pan2020band}. At $g=0$, it is valley polarized but now the holes populate the valley whose spin is anti-aligned to the direction of the magnetic field. For a given $B$, it can be thought of as a Landau quantized time reversed partner of the QAH discussed above. Its (large) gap also does not vanish in the $B\rightarrow 0$ limit, see Fig.\ref{fig:my2}(d). If it were stable, the valley polarization would reverse at a fixed $B$ upon changing the filling from $-1-\frac{|\phi|}{\phi_0}$ to $-1+\frac{|\phi|}{\phi_0}$, similar to the observations reported in Ref.\cite{Polshyn2020} in graphene heterostructures. Such a state would appear as a prominent Chern insulator, inconsistent with the experiments\cite{cai2023signatures,zeng2023thermodynamic,park2023observation, xu2023observation}. At $g\gtrsim2$, however, the ground state changes and the holes populate the valley with spin aligned to the direction of the magnetic field for a range of $B$ that expands to lower $|B|$ as $g$ increases (Fig.\ref{fig:my2}(h)). It can be thought of as an integer quantum Hall (IQH) state, populating the lowest pair of Landau levels (LLs) of the Chern paraelectric excitation spectrum at $t=-\text{sign}(B)$ with electrons, as seen in the Fig.\ref{fig:my2}(c). Because of the sizable band mass, its gap is small and vanishing as $B\rightarrow 0$, consistent with the experimental absence of a prominent Chern insulator at $t=+\text{sign}(B)$ and presence at $t=-\text{sign}(B)$. As detailed below, we estimate the orbital Zeeman contribution to $g$ in tMoTe$_2$ to be $3.23$ and of the same sign as the spin contribution $2$, placing us safely in the $g\gtrsim 2$ regime.

Perpendicular electric field, $D$, has also proved to be a powerful tool to probe and manipulate the nature of correlated states in tMoTe$_2$\cite{park2023observation, zeng2023thermodynamic, cai2023signatures, xu2023observation}. In the experiments at $B=0$, the spin polarized QAH state transitions into a trivial insulator above a critical $D_*$\cite{park2023observation, zeng2023thermodynamic, cai2023signatures, xu2023observation}.
The proximate trivial state is argued to be spin, and therefore valley, unpolarized in the Ref.~\cite{zeng2023thermodynamic} and spin, and therefore valley, polarized in the Ref.~\cite{xu2023observation}. Such topological transition is also seen in the Hartree-Fock calculation at $B=0$\cite{wang2023topology,li2024electrically}, and the spin structure of the trivial state depends on the value of $\epsilon$. For $\epsilon>\epsilon'\approx 19$ the transition is directly into a spin unpolarized intervalley coherent state (IVC) and for $\epsilon_*\approx 15.4<\epsilon<\epsilon'$ the proximate trivial state is valley polarized (VP). This can be seen in the $B=0$ plane of Fig.\ref{fig:my4} where we vary the $D$-induced potential difference between two layers $u_D$. Intriguingly, at $B\neq0$ and $t=-\text{sign}(B)$ a sharp change of compressibility can be seen in the experiment at the same $D_*$ as the transition at $B=0$ (see Extended Data Fig.10 of Ref.~\cite{zeng2023thermodynamic}). Our calculation at $B\neq0$ captures this effect. As shown in the Fig.\ref{fig:my3}(a), at $\epsilon=20$ chosen to recover the phenomenology of Ref.\cite{zeng2023thermodynamic} at $B=0$, we observe a crossover at $D_*$ between a Chern paraelectric with a large gap and an ostensibly IQH state of a Landau quantized trivial insulator with one LL depopulated by electrons and a small gap. We refer to the latter as a QH ferroelectric because at $u_D=0$ it breaks $C_{2y}\mathcal{T}$. Both states are spin polarized. The results of a separate Hartree-Fock calculation at $t=0$ shown in the Fig.\ref{fig:my3}(b) confirm the prominent trivial insulator with a large gap.
Thus, the incompressible states can appear simultaneously at $\nu=-1$ and $\nu=-1 - \frac{|\phi|}{\phi_0}$ for a range of $D$ near $D_*$.

In order to explore the robustness of the above results, as well as the tunability of the correlated phases, we perform the Hartree-Fock calculations at $s=-1$ and $t=-\text{sign}(B)$ at varying $\epsilon$, $u_D$ and $B$. The resulting tentative phase diagram is shown in the Fig.\ref{fig:my4}. At $D=0$ the first order topological transition at $\epsilon_*$ extends to $B\neq0$ despite separating states with the same Chern number $t$ at $B\neq0$ because the QH ferroelectric on one side breaks $C_{2y}\mathcal{T}$ and the Chern paraelectric on the other does not. This first order phase transition moves towards stronger interaction as $B$ increases, favoring the Chern paraelectric i.e. the descendant of the QAH state. The transition must also extend to small non-zero $D$ despite the explicit breaking of $C_{2y}\mathcal{T}$ by $D$ because it starts out first order at $D=0$. Indeed, Fig.\ref{fig:my3}(c) shows the jump in the layer polarization at a fixed $B$ persisting to non-zero $u_D$, with $D$ favoring the ferroelectric and the first order transition terminating at a critical endpoint. As we show below, we can describe this using a simple Landau theory for an Ising ferroelectric with a negative quartic coupling. Upon varying $B$ the critical endpoint extends to the red curve shown in the Fig.\ref{fig:my4}. Interestingly, the $D$ induced crossover shown in the Fig.\ref{fig:my3}(a) appears at $B$ and $\epsilon$ notably separated from the critical endpoint. We understand it as crossing the Widom line, defined as the peak in the layer polarizability, extending beyond the critical endpoint as shown in the Figs.\ref{fig:my3}(d) and \ref{fig:my4}.  As IVC is strongly suppressed by the $B$-field, Chern paraelectric and QH ferroelectric are the main phases in the Fig.\ref{fig:my4}.
Below we discuss our formalism and computational scheme that lead to the above results. We note in passing that at the twist angle $3.89^\circ$, $\phi=\phi_0$ at $B\simeq176$T, so to get below $9$T as in the experiments we need $\phi/\phi_0\lesssim1/20$. Reaching such a small flux efficiently in our interacting calculation is enabled by a recently developed technique\cite{wang2022narrow,wang2023revisiting,supplementary} utilizing the $B=0$ hybrid Wannier states to construct the basis at $B\neq0$; some of our calculations go down to $\phi /\phi_0 = 1/47$.

\begin{figure}
\centering
\includegraphics[width=\linewidth]{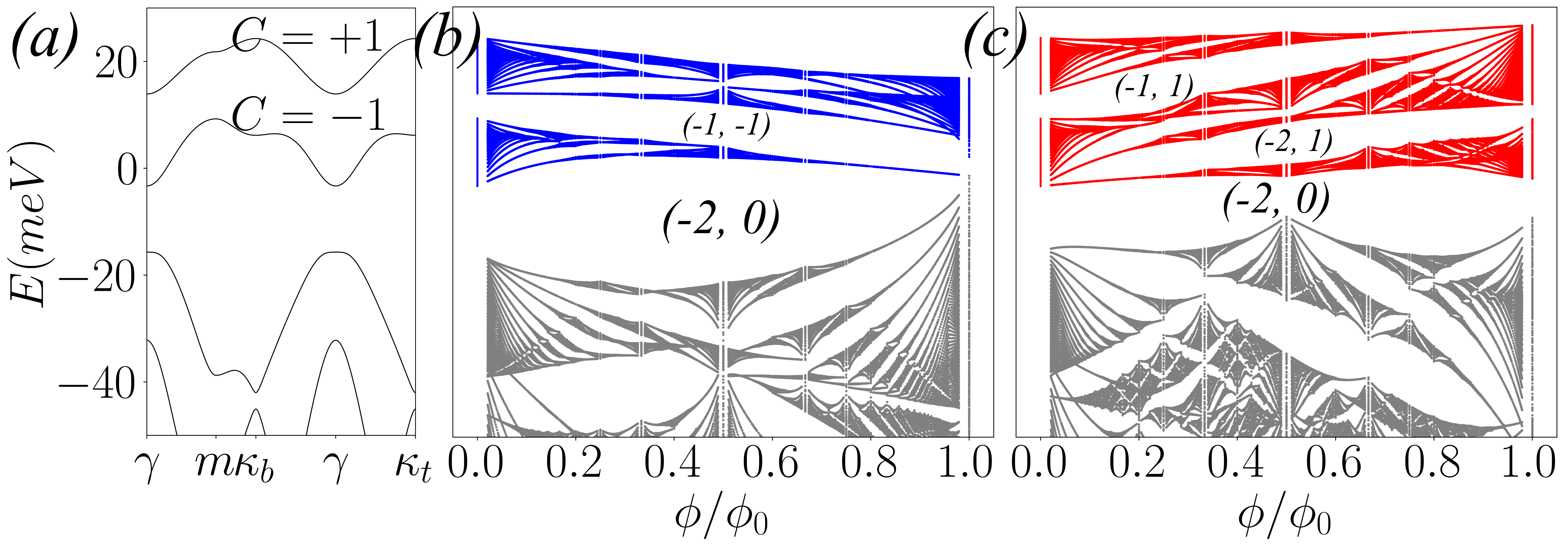}
\caption{(a) Band structure at $B=0$ in the valley $\bK$ for tMoTe$_2$ at $\theta = 3.89^{\circ}$.(b,c) Hofstader spectra at $\bK$ and $\bK'$, omitting Zeeman effect.}
\label{fig:my1}
\end{figure}
\begin{figure*}
\centering
\includegraphics[width=\linewidth]{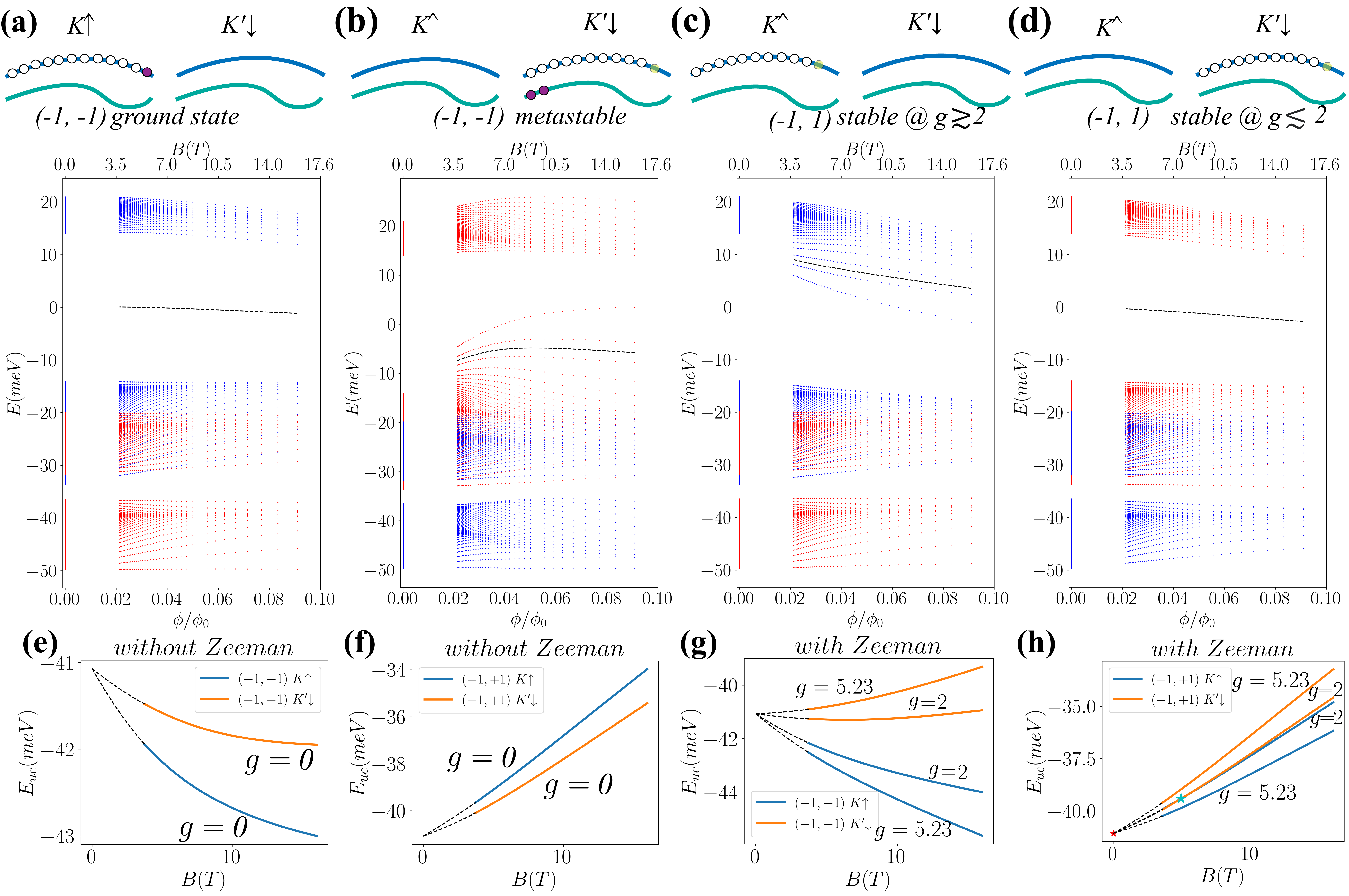}
\caption{(a-d)Hartree Fock energy spectrum with varying magnetic field at $\nu = -1 - \frac{|\phi|}{\phi_0}$ and $\nu = -1 +\frac{|\phi|}{\phi_0}$. Electrons occupy all eigenstates below black dashed line (guide to eye). Blue (red) points represent eigenstates in $\bK$($\bK'$) valley. (e,f)Total energy per unit cell versus magnetic field. Dashed lines are obtained from quadratic interpolation using calculated energy per unit cell at $B=0$. (g,h)Energy per unit cell with non-zero $g$ factor.}
\label{fig:my2}
\end{figure*}

\PRLsec{Continuum model} The moir\'e bands of experimental relevance originate from the spin-valley locked valence bands of monolayer MoTe$_2$\cite{xiao2012coupled}.
The relations in Eq.~(\ref{eq:Streda}) and Eq.~(\ref{eq:Diophantine}), as well as the stated experimental results, hold in any right handed coordinate system (even though $\sigma_{xy}$, and therefore $t$, changes sign between two coordinate system choices related by a $180^\circ$ rotation about $x$ or $y$ axis). We are therefore free to choose any of the two out-of-plane directions as $\hat{\bf z}$ and we made the common choice \cite{xiao2012coupled} of aligning it with the spin angular momentum direction of the valley $\bK$.
Then, in valley $\bK$ and for spin up, the non-interacting moir\'e band structure can be described by a continuum electronic model \cite{wu2019topological}:
\begin{equation} \label{eq:nonintHam}
\begin{split}
    & {H}_{\bK} = \int \mathrm{d}^2 \br
    \begin{pmatrix}
        \psi^{\dagger}_{b \bK}(\br) & \psi^\dagger_{t \bK}(\br)
    \end{pmatrix} \times \\
    & \begin{pmatrix}
    {h}_{b}({\bp}) + \Delta_{b}(\br) + \frac{u_D}{2} & \Delta_{T}(\br) \\
    \Delta_{T}^{*}(\br) & {h}_{t}({\bp}) + \Delta_{t}(\br) - \frac{u_D}{2}
    \end{pmatrix}
    \begin{pmatrix}
        \psi_{b \bK}(\br)\\ \psi_{t \bK}(\br)
    \end{pmatrix}.
\end{split}
\end{equation}
Here $\psi^{\dagger}_{l \bK}(\br)$ creates an electron in layer $l=b, t$ , in valley $\bK$, and at position $\br$. ${h}_{l}({\bp}) = -({\mathbf{p}}- \hbar \kappa_{l})^2/2m^*$ is kinetic energy, with ${\bp}=-i\hbar \nabla_{\br}$ and effective mass $m^{*}$. $\kappa_{l}$ points sit at the corners of the hexagonal moir\'e Brillouin zone,
we describe the effect of $D$ by adding a potential difference $u_D$ between two layers, $\Delta_{b(t)}(\br) = 2 V \sum_{j = 1, 3, 5} \cos(\bG_j \cdot \br \pm \phi)$ is the intralayer potential for bottom(top) layer, and $\Delta_T(\br) = w(1 + e^{-i \bG_2 \cdot \br} + e^{-i \bG_3 \cdot \br})$ is interlayer tunneling potential. The reciprocal lattice vector $\bG_j$ is obtained by counterclock-wise rotation of $\bG_1= 4\pi/(\sqrt{3}a_M) \hat{\mathbf{y}}$ by angle $(j-1)\pi / 3$. $a_M = a_0 / \theta[\text{rad}]$ is moir\'e superlattice constant, where $a_0$ is the lattice constant of monolayer MoTe$_2$. We choose parameters $(a_0,m^*, V, \phi, \omega) = (0.355 {\rm nm}, 0.62 m_e, 17.0 {\rm meV},+107.7^{\circ},-16.0 {\rm meV})$ at twist angle $\theta = 3.89^{\circ}$ \cite{wang2023topology}. The non-interacting model in the $\bK'$ valley and for spin down is related by time reversal symmetry $\mathcal{T}$, which takes the complex conjugation of the $2\times2$ Hamiltonian in Eq.~(\ref{eq:nonintHam}).

In an external magnetic field, we replace ${\bp}\rightarrow {\bp} + e\bA$, where $\nabla\times\bA=B\hat{\bf z}$, and add Zeeman coupling $g\mu_BB \tau_0 s_z$, where $\tau_0$ is the identity Pauli matrix acting on layer degrees of freedom, and $s_z$ is the spin Pauli matrix. At $B=0$, the Hamiltonian is invariant under three-fold out-of-plane rotation $C_{3z}$, in-plane 2-fold rotation symmetry about the $y$-axis $C_{2y}$ and the time reversal symmetry $\mathcal{T}$. At this order the continuum Hamiltonian also has a pseudo-inversion symmetry\cite{yu2024fractional, jia2023moir} $\mathcal{I}$ given by $\mathbf{r} \rightarrow -\mathbf{r}$ in each valley and flipping the layers with $\tau_1$. While both $C_{2y}$ and $\mathcal{T}$ are broken at $B \neq 0 $, their product $C_{2y}\mathcal{T}$ is preserved, as are $C_{3z}$ and $\mathcal{I}$.

\begin{figure}
\centering
\includegraphics[width=\linewidth]{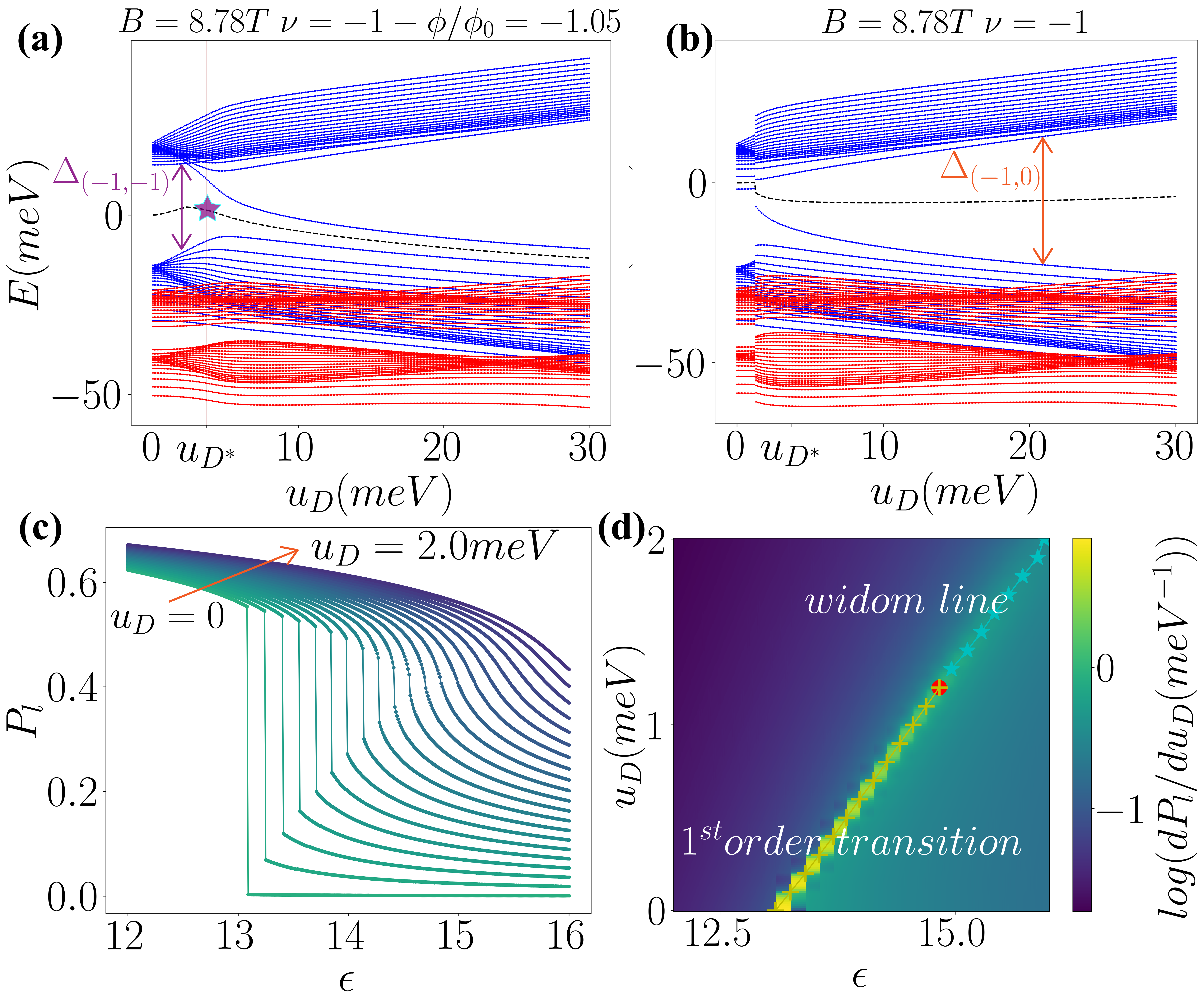}
\caption{(a)Hartree Fock spectrum at $\epsilon=20$ with varying $u_D$ at $\nu = -1 - \frac{|\phi|}{\phi_0}$ and (b) $\nu = -1$. Topological phase transition at $B=0$ is at $u_{D^{*}}$. (c,d) Layer polarization $P_l$ and polarizability with varying $u_D$ and $\epsilon$ at $B=8.78$T.}
\label{fig:my3}
\end{figure}
Fig.~\ref{fig:my1}(a) shows the non-interacting band structure at $B=0$ in the valley $\bK$ where the two uppermost bands carry Chern number $+1$ and $-1$. The band dispersion and Chern numbers in the opposite valley $\bK'$ are related by $\mathcal{T}$.
For $B>0$ the corresponding Hofstadter spectra for the valleys $\bK$ and $\bK'$, together with the $(s,t)$ labels of the gaps, are shown in Fig.~\ref{fig:my1}(b) and (c) respectively. As can be seen by the evolution of the magnetic subbands, a Chern band gains or loses states per unit cell according to the relation $1+t\phi/\phi_0$.
To study correlated phenomena, we project the Coulomb interaction onto the upper two Chern $\pm1$ bands of both valleys if at $B=0$, i.e. two bands per valley, or the magnetic subbands emanating from these Chern bands if at $B\neq0$,
\begin{equation} \label{eq:myInteraction}
    {H}_V = \frac{e^2}{4\epsilon\epsilon_0 A}\sum_{\bq\neq0}\frac{1}{ |\bq|}\tanh\left(|\bq|d\right) :\mathcal{P}{\rho}_{\bq}{\rho}_{-\bq} \mathcal{P}:.
\end{equation}
Here $\rho_{\bq} = \sum_{\eta, l} \int \mathrm{d}^2 \br e^{-i \bq \cdot \br} \psi_{l\eta}^{\dagger}(\br) \psi_{l\eta }(\br)$ is the Fourier transform of the electron density operator, with $\eta=\bK$ or $\bK'$, and total area $A$. In this work we consider a dual gate screened Coulomb interaction, $d$ being the distance between each gate and tMoTe$_2$, taken to be $30$nm in our calculations,  with the small distance between the two MoTe$_2$ layers neglected. Symbol $::$ denotes operator normal ordering.
$\mathcal{P}$ projects onto the mentioned Hilbert subspace which can be generated at $B\neq0$ by solving the non-interacting problem\cite{wang2023revisiting, wang2022narrow, foutty2023mapping}. At the $B$ values of interest here, expanding $\mathcal{P}$ in the LL basis proves to be computationally expensive when dealing with interactions. Instead, we construct the basis states using the $B=0$ hybrid Wannier states method following Ref.~\cite{wang2023revisiting, supplementary}.

\PRLsec{Streda line near $\nu = -1$ and Zeeman effect} The $B>0$ energy spectrum of candidate Chern states along the $(-1, \mp1)$ lines is shown in Fig.~\ref{fig:my2}(a-d). We present their total energy per unit cell as a function of $B$ at varying Zeeman $g$ factors in Fig.~\ref{fig:my2}(e-h). The orbital contribution to the Zeeman effect stems from $d$-orbital nature of the states in each valley\cite{xiao2012coupled}. The two-band $\bk\cdot\bp$ Hamiltonian describing the large semiconductor gap in MoTe$_2$ monolayer has the form of a massive 2+1D Dirac Hamiltonian\cite{xiao2012coupled}, whose minimal coupling to a low $B$-field results in an additional $\frac{1}{2}\hbar\omega_c$ energy shift compared to a Schrodinger Hamiltonian. Since the Dirac mass effectively flips sign between different valleys, this orbital shift also flips sign; it adds to the usual spin Zeeman factor of $2$ (see Ref.~\cite{supplementary} for details). Using the known effective mass of MoTe$_2$ we find it to be $3.23$ and thus the total $g=5.23$.
Using the computed ground state energy of the QAH state at $B=0$, we smoothly interpolate between the calculated energies at $B\neq0$ and $B=0$ at $g=0$ using at most a second order polynomial in $B$ as shown by the dashed lines in
Fig.\ref{fig:my2}(e,f).
Because the Chern paraelectrics are fully valley polarized, their Zeeman energy can be obtained analytically. We can therefore find the critical $B$ below which the valley switches for Chern paralectrics. As seen in the Fig.\ref{fig:my2}(g,h)  the crossing only occurs for the $(-1, 1)$ line. While at $g=2$ it occurs at $5.01$T (cyan star), at the more realistic $g = 5.23$ it is $\sim178$mT (red star). Thus any putative density induced valley switching would occur in the regime where the effects of quenched disorder dominate in realistic devices\cite{park2023observation, xu2023observation}, likely modifying the clean limit energetics obtained here

\begin{figure}
\centering
\includegraphics[width=\linewidth]{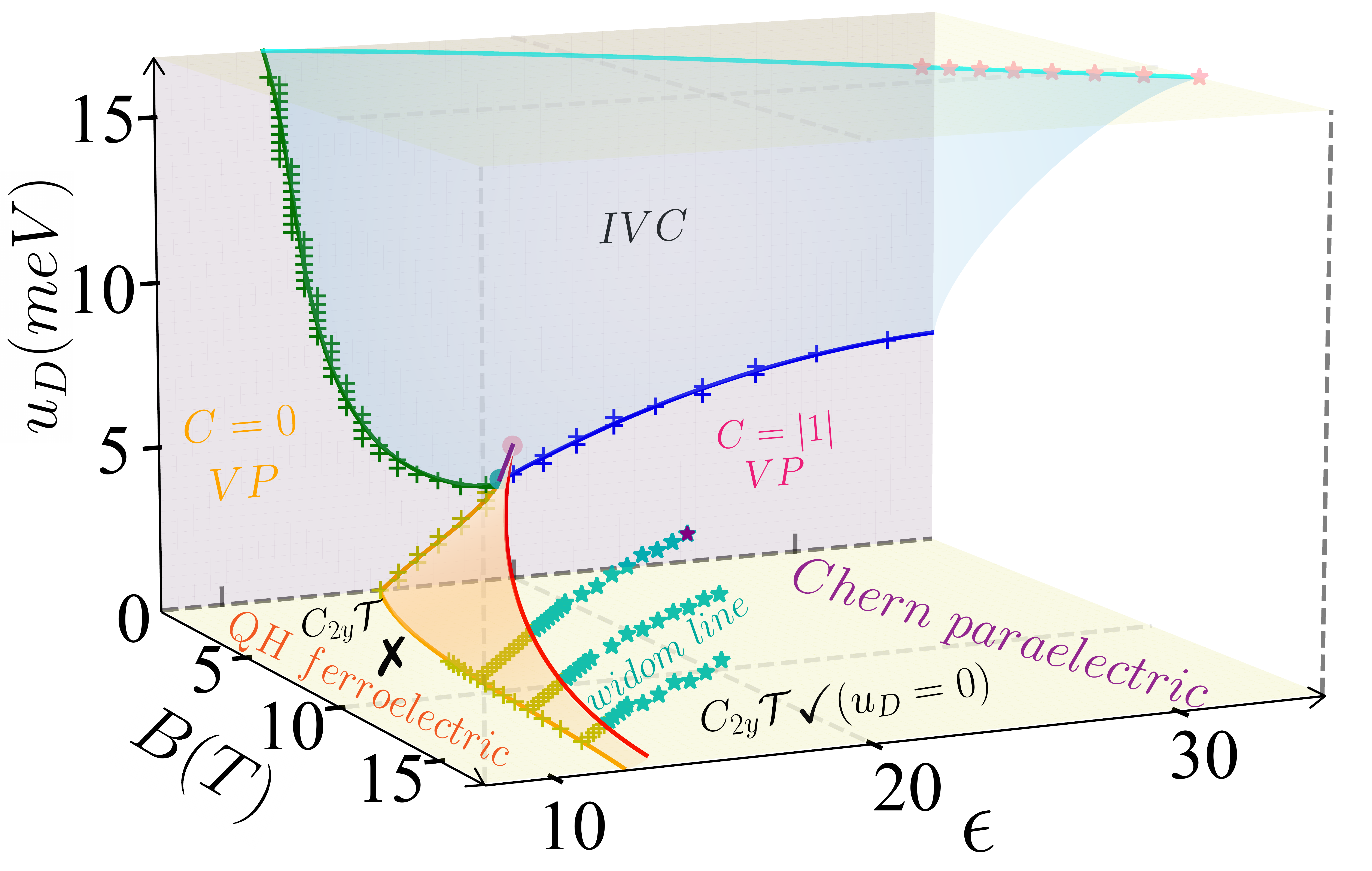}
\caption{Tentative phase diagram of tMoTe$_2$ at $\nu = -1 -\frac{|\phi|}{\phi_0}$ and $3.89^\circ$ twist. The blue surface at $B \neq 0$ denotes the boundary between IVC state and two valley $U(1)$ preserving states, Chern paraelectric and QH ferroelectric.}
\label{fig:my4}
\end{figure}

\PRLsec{Effects due to $D$-field}
Energy spectrum induced by the perpendicular electric field $D$ at $\phi / \phi_0 = 1/ 20$, corresponding to $B = 8.78T$, is shown in Fig.~\ref{fig:my3}(a) along the $(-1,-1)$ line.
The lowest unoccupied LL can be seen to detach from the unoccupied LL group as $D$ increases and to closely approach the highest occupied LL. This is similar to the magnetic subband spectrum of an electron confined to move in a plane in a periodic potential and an out-of-plane uniform magnetic flux slightly smaller than $\phi_0$ shown in the Fig. 7 of Ref.\cite{tevsanovic1989hall} as the strength of the periodic potential increases and drives a transition from a Chern insulator (a broadened LL) and a trivial insulator at their $\phi=\phi_0$. As shown in the Fig.~\ref{fig:my3}(b) for $(-1,0)$, a prominent trivial insulator is indeed stabilized in our Hartree-Fock calculation as well.

The evolution of the layer polarization $P_l=\frac{A_{uc}}{A}\sum_{\eta} \int \mathrm{d}^2 \br \langle \psi_{t\eta}^{\dagger}(\br) \psi_{t\eta}(\br)-\psi_{b\eta}^{\dagger}(\br)\psi_{b\eta}(\br) \rangle$
and polarizability $dP_l/du_D$ at $(-1,-1)$ with varying $\epsilon$ and $D$, at a fixed $B = 8.78T$, are shown in Fig.~\ref{fig:my3}(c,d).
The main features can be captured by a simple Landau theory for an Ising order parameter, odd under $C_{2y}\mathcal{T}$,
\begin{equation} \label{eq:myLandau}
\begin{gathered}
    \delta E = -\gamma u_D P_{l} + \frac{a(\epsilon)}{2}  P_l^2 - \frac{b}{4}  P_l^4 + \frac{1}{6} P_l^6.
\end{gathered}
\end{equation}
Here $\gamma$ and $b$ are positive constants, and the prefactor of $P_l^6$ was used to rescale the energy units. Decreasing $a$ with decreasing $\epsilon$ causes the spontaneous symmetry breaking at $u_D\propto D =0$ via a 1$^{st}$ order phase transition due to the negative quartic term. At the critical endpoint, the local minima merge (see Ref.\cite{supplementary}).

\PRLsec{Phase diagram} The tentative 3D phase diagram along the $(-1, -1)$ line is shown in the Fig.~\ref{fig:my4}. Markers correspond to parameters at which the transition, or a crossover in the case of Widom line, was calculated within Hartree Fock. The solid lines and surfaces are interpolated based on physical arguments. At $u_D\propto D=0$ the Chern paraelectric and the quantum Hall ferroelectric are distinguished by $P_l$, a Landau order parameter, and so must be separated by a phase transition. At $u_D\neq0$ however, it is possible to transform one to another without encountering a phase transition by avoiding the (orange) surface of 1$^{st}$ order phase transitions terminating in the (red) critical endpoint curve. The IVC is a sharply defined phase at any $B$ and $D$ because it breaks valley $U(1)$ symmetry, preserved by $B$ and $D$ while the other phase(s) do not break it.
It is suppressed by $B$ even at $g=0$, and further suppressed by the Zeeman effect. Therefore, while all the lines are computed at $g=5.23$, the (cyan) phase boundary at $u_D=16.8$meV is extrapolated from the $g=2$ markers; it moves closer to the $B=0$ plane at higher $g$ (see Ref.\cite{supplementary}).

\PRLsec{Discussion} Our calculation explains the presence of the $(-1,-1)$ line and the absence of the $(-1, +1)$ line in the experiments\cite{cai2023signatures, xu2023observation, zeng2023thermodynamic, park2023observation}. It also demonstrates the delicate balance between the energies of competing Chern states at $(-1,+1)$ polarized to opposite valleys, with the orbital Zeeman contribution ultimately deciding on the ground state with the small gap. Recent calculations\cite{zhang2023polarization, mao2023lattice, jia2023moir} pointed out a rich evolution of the Chern bands and a lattice relaxation playing increasingly important role at low twist angles. While the Zeeman contribution is expected to be the same in the low angle regime, the energetic differences due to the magnetic subbands shown in our Fig.\ref{fig:my2}(e,f) will change. It would be interesting to explore the possibility of density induced switching of the Chern number at different angles or different homobilayers using the framework developed here.

\PRLsec{Acknowledgement} We thank B. Andrei Bernevig, Fengcheng Wu, Di Xiao, Jian Kang, Eslam Khalaf, Jiabin Yu, Daniel Parker, Kin Fai Mak, Jie Shan, and Xiaodong Xu for helpful discussions. M.W. sincerely thanks Taige Wang for fruitful discussions and ongoing collaborations. 
X.W. acknowledges financial support from the National High Magnetic Field Laboratory through NSF Grant No.~DMR-2128556 and the State of Florida.
O.V. was funded by the Gordon and Betty Moore Foundation’s EPiQS Initiative Grant GBMF11070. Part of the numerical computations were
carried out on resources provided by the Planck clusters at
Florida State University. M.W. acknowledge the Beijing Paratera Co., Ltd. for providing HPC resources that have contributed to the research results reported within this paper.

\bibliography{arXiv_submission}

\clearpage

\begin{widetext}

\begin{center}
  \textbf{\large Supplemental Material for ``Interacting phase diagram of twisted MoTe$_2$ in magnetic field"}
\end{center}

\setcounter{equation}{0}
\setcounter{section}{0}
\setcounter{figure}{0}
\renewcommand{\theequation}{S\arabic{equation}}
\setcounter{table}{0}

\tableofcontents

\section{Moir\'e geometry}

Moir\'e primitive and reciprocal primitive lattice vectors are shown in Fig.~\ref{fig:myMBZschematic}, and given explicitly as follows:
\begin{equation} \label{eq:MYaAndG}
\ba_1 = a_M\left(\frac{\sqrt{3}}{2}, -\frac{1}{2}\right), \ba_2
 = a_M(0, 1),\
\textbf{G}_1 = \frac{4\pi}{\sqrt{3}a_M}(1, 0), \textbf{G}_2 = \frac{4\pi}{\sqrt{3}a_M}\left(\frac{1}{2}, \frac{\sqrt{3}}{2}\right),
\end{equation}
where the moir\'e period $a_M = {a_0}/{\theta}$. At $B = 0$, we take the rhombus Brillouin zone with orange boundary lines in Fig.~\ref{fig:myMBZschematic}, with wavevector defined as $\bk=k_1\bG_1+k_2\bG_2$  where $k_{i=1,2}\in[0,1)$. At $B\neq 0$, we use the Landau gauge $\bA = (0, Bx)$ and work at rational magnetic flux per unit cell ${\phi}/{\phi_0} = {p}/{q}$, where $\{p,q\}$ are coprime integers and $\phi_0=h/e$ is the flux quantum. Hence, magnetic translation operators are
\begin{equation} \label{eq:myMagTa1a2}
\hat{t}(\ba_1) = e^{-i \bq_\phi \cdot \br} \hat{T}(\ba_1) = e^{-i\frac{2\pi}{a_M} \frac{p}{q} \textbf{y}} \hat{T}(\ba_1),\hat{t}(\ba_2) = \hat{T}(\ba_2)
\end{equation}
with $\hat{T}(\ba_{\{ 1, 2\}})$ translation operators at $B = 0$(see, for example, Ref.\cite{wang2023revisiting}). Since $[ \hat{t}(\ba_1), \hat{t}^q(\ba_2)] = 0$, we can construct magnetic moir\'e Brillouin zone as $\{k_1 \bG_1 + k_2 \bG_2 | 0<k_1<1, 0<k_2<\frac{1}{q} \}$, whose boundary lines are shown as purple at $\phi / \phi_0 = 1 / 11$. At $\phi/\phi_0\ll 1$, we find it sufficient to perform Hartree-Fock calculations with a discretized mesh given by the cyan and brown points. Note if the Hartree-Fock ground state preserves $\hat{t}(\ba_2)$, the calculations can be further simplified by constructing the Hartree-Fock Hamiltonian at the cyan point only\cite{wang2023revisiting, wang2022narrow}. When constructing hybrid Wannier states at $B = 0$, we consider real space as $N_1 \ba_1 \times N_2 \ba_2$, which makes $\frac{A_{uc}}{A} = \frac{1}{N_1 N_2}$ and discrete $\bk$ points $\bk = \frac{n_1}{N_1} \bG_1 + \frac{n_2}{N_2} \bG_2 = k_1 \bG_1 + k_2 \bG_2$ with integers $n_1$ and $n_2$.

\begin{figure}
\centering
\includegraphics[width=0.8\linewidth]{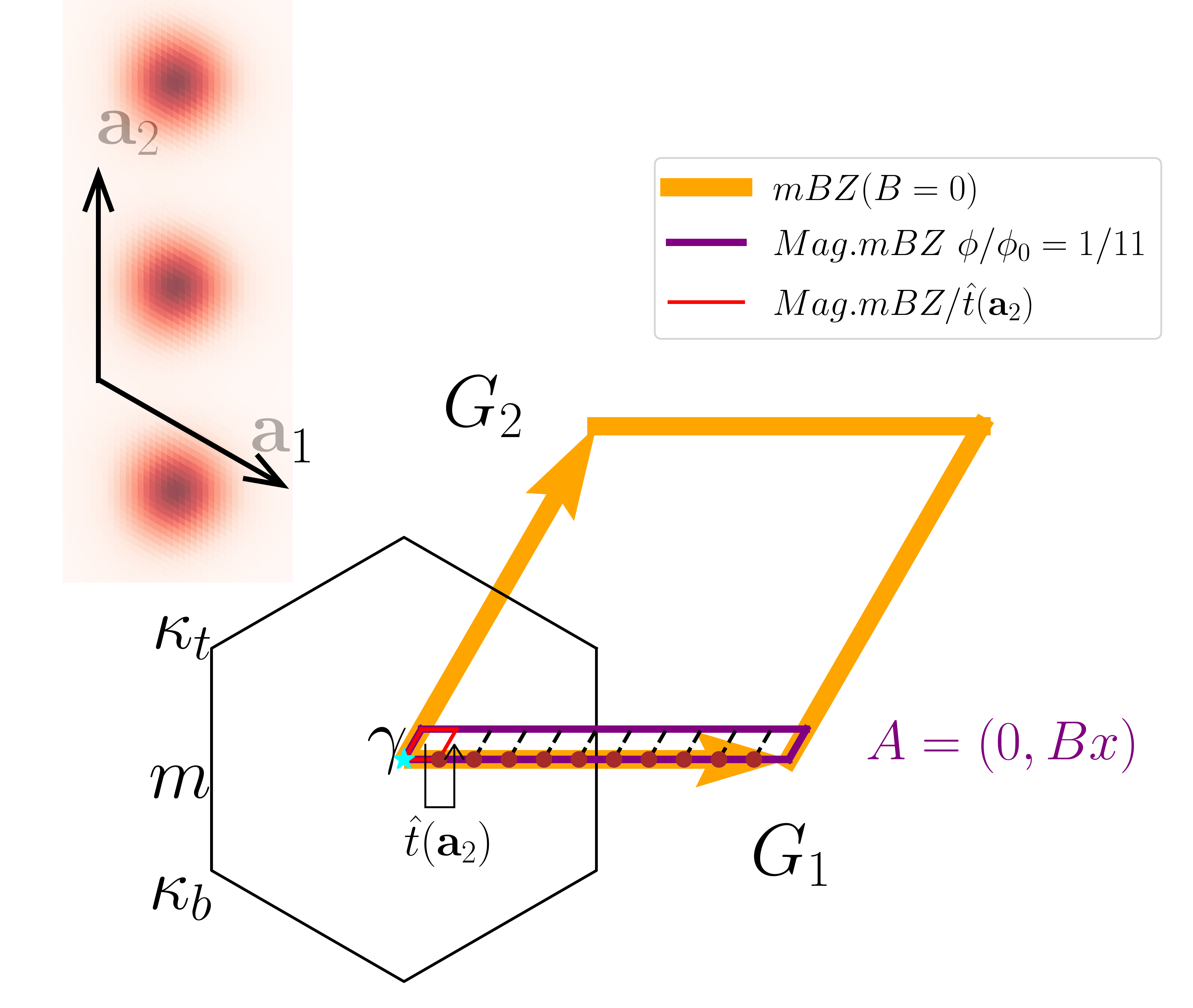}
\caption{Schematics of moir\'e Brillouin zone at $B = 0$ and $B \neq 0$. Sample points are also shown in the figure. Inset plot shows probability density of hybrid Wannier states in tMoTe$_2$.}
\label{fig:myMBZschematic}
\end{figure}

\section{Orbital contribution to $g$ factor}
Correct band basis and fitting parameters in MoTe$_2$ are crucial to estimate the orbital contribution to Zeeman effect. The Ref.~\cite{xiao2012coupled} is based on monolayer calculation and Ref.~\cite{wu2019topological} on aligned bilayer; their band basis of MoTe$_2$ near $\bK$ and $\bK'$ valley are the same. Namely, conduction bands in both valleys are described by $d_{z^2}$ orbital, valence band in $\bK$ is by $d_{x^2 - y^2} + id_{xy}$ and valence band in $\bK'$ by $d_{x^2 - y^2} -id_{xy}$. The effective $\bk \cdot \bp$ models (for one layer) differ only by $\sigma_0$ term in valence-conduction $2 \times 2$ subspace, which simply shifts total energy and will not affect effective mass in each valley.
We therefore choose to start with massive Dirac Hamiltonian in $\bK$ valley\cite{wu2019topological}
\begin{equation} \label{eq:myKDirac}
    h_l^{\bK}(\bk, d_0) = \begin{pmatrix}
    \Delta_g + \Delta_{l, c}(d_0) & \hbar v_{F} (k_x - i k_y) \\
    \hbar v_{F} (k_x + i k_y) & \Delta_{l, v}(d_0)
    \end{pmatrix}
\end{equation}
here $c$ denotes conduction bands above semiconductor gap $\Delta_g$, and $v$ denotes valence bands below the gap. Since $\Delta_g$($\approx 2eV$) is much larger than other energy scales, we could write eigen-equations for valence band wavefuction $\ket{\psi_v}$ as
\begin{equation}
\begin{split}
    &\hbar v_{F}(k_x + ik_y) \frac{-\hbar v_{F}(k_x - i k_y)}{\Delta_g + \Delta_{l, c}(d_0) - E} \ket{\psi_v} + \Delta_{l, v}(d_0) \ket{\psi_v} \\
    &\approx \frac{-\hbar^2 v_F^2}{\Delta_g} (k_x + i k_y)(k_x - i k_y) \ket{\psi_v} + \Delta_{l, v}(d_0) \ket{\psi_v} \\
    &= (-\frac{p^2}{2m^{*}} + \Delta_{l, v}(d_0)) \ket{\psi_v} = E\ket{\psi_v}, \\
\end{split}
\end{equation}
where $m^{*} =\frac{\Delta_g}{2 v_F^2} \approx 0.62 m_e$ is the effective mass for valence band electron \cite{wu2019topological, xiao2012coupled}. Within this approximation, the Dirac equation gives an approximately parabolic kinetic energy for valence band electrons.

Now consider adding magnetic field through minimal coupling. Note the original form of kinetic energy is $\frac{-1}{2m^{*}}(p_x + ip_y)(p_x - ip_y)$ for $\bK$ valley electron. Minimal coupling within the original form is not equal to the minimal coupling within $\frac{-p^2}{2 m^*}$, because the momentum operator does not commute with the vector potential $\bA$,
\begin{equation}
\begin{split}
    -\frac{1}{2m^*}(p_x + i p_y)(p_x - i p_y)&\rightarrow
  -\frac{1}{2m^*}(p_x +eA_x + ip_y + ieA_y)(p_x + eA_x - ip_y - ieA_y)\\
    &=-\frac{1}{2m^*}((p_x+eA_x)^2 + (p_y + eA_y)^2 + i[p_y + eA_y, p_x + eA_x])\\
    &=-\frac{1}{2m^*}((p_x+eA_x)^2 + (p_y + eA_y)^2 + ie([p_y, A_x] + [A_y, p_x]))\\
    &=-\frac{1}{2m^*}((p_x + eA_x)^2 + (p_y+eA_y)^2) - \frac{1}{2m^*}ie(-i\hbar \partial_y A_x + i\hbar \partial_x A_y)\\
    &=-\frac{1}{2m^*}((p_x + eA_x)^2 + (p_y+eA_y)^2) -\frac{1}{2m^*}ie(+i\hbar)(\nabla \times A)\\
    &=-\frac{1}{2m^*}((p_x + eA_x)^2 + (p_y+eA_y)^2) + \frac{e}{2m^*}\hbar B.
\end{split}
\end{equation}
The first term in the last line is already included by replacing $\bp \rightarrow \bp + e\bA$ in the continuum model at $B = 0$ presented in the main text, and the second term can be understood as the orbital contribution to the Zeeman energy shift. Indeed, comparing to the definition of the $g$ factor $H_{Zeeman} = -\mu \cdot B = \frac{e}{2m_e}g \frac{\hbar}{2}B$ we get
\begin{equation}
g_{orbit} = 2\frac{m_e}{m^*} = 2 \times \frac{1}{0.62} \approx 3.23.
\end{equation}

The massive Dirac Hamiltonian in the $\bK'$ valley is different from Eq.~\ref{eq:myKDirac}, since the valence band basis changes\cite{xiao2012coupled}:
\begin{equation} \label{eq:myKprimeDirac}
    h_l^{\bK'}(\bk, d_0) = \begin{pmatrix}
    \Delta_g + \Delta_{l, c}(d_0) & \hbar v_{F} (-k_x - i k_y) \\
    \hbar v_{F} (-k_x + i k_y) & \Delta_{l, v}(d_0)
    \end{pmatrix}.
\end{equation}
Therefore for $\bK'$ valley electrons, kinetic energy reads $\frac{-1}{2m^{*}}(p_x - ip_y)(p_x + ip_y)$, and hence additional energy shift is $-\frac{e}{2m^*}\hbar B$. The opposite sign of this orbital contribution is again aligned with the opposite spin orientation in the valley $\bK'$ due to spin-valley locking, making the total $g$ factor in both valleys approximately $5.23$.

\section{Landau Theory of transition between Chern paraelectric and QH ferroelectric}

In order to further explore the first order phase transition and the related crossover physics, we write a phenomenological Landau theory and show that it can capture the results of Hartree Fock calculation presented in the main text.
At $u_D=0$ the Hamiltonian is invariant under $C_{2y}\mathcal{T}$ which is spontaneously broken in the quantum Hall ferroelectric phase and preserved in the Chern paraelectric phase. The layer polarization $P_l$, which changes sign under $C_{2y}\mathcal{T}$, can therefore serve as a Landau order parameter. The Landau free energy must be invariant under $C_{2y}\mathcal{T}$ and therefore can only contain even powers of $P_l$ at $u_D=0$.
In order to obtain the first order transition at $\epsilon_*$, one needs a negative prefactor of the $P_l^4$ term (and of course a positive prefactor of the $P_l^6$ term in order to stabilize the free energy).  Non-zero $u_D$ then couples directly to $P_l$ at linear order. Therefore,
\begin{equation} \label{eq:SMLandau}
\begin{gathered}
    \delta E = -\gamma u_D P_{l} + \frac{1}{2} a(\epsilon) P_l^2 - \frac{1}{4} b P_l^4 + \frac{1}{6} P_l^6.
\end{gathered}
\end{equation}
Here the prefactor of $P_l^6$ term is set to $1/6$ for convenience which can always be achieved by rescaling $\delta E$. Constant $\gamma$ is added to set the minimum position of $\delta E$ at $0<P_l<1$. Near the phase boundary, the prefactor of $P_l^2$ term is treated as a function of dielectric constant $\epsilon$, while the prefactor $b$ is approximated as a constant. Note $b < 0$ will turn transition into a continuous one, hence is not relavant to the phase transition studied in the main text.

At $u_D=0$ the free energy has a minimum at $P_l=0$ and, depending on the value of $a(\epsilon)$, it may have another pair of equal and opposite minima at non-zero $P_l$. At $a(\epsilon)=3b^2/16$ the minima are degenerate and this corresponds to the first order phase transition at $u_D=0$.

At small non-zero $u_D$ one of the minima shifts away from $P_l=0$, the degeneracy between the other two minima is lifted, and the transition remains first order. At very large $u_D$ there is a single minimum. To find the critical endpoint, we analyze the extrema of $\delta E$:
\begin{equation} \label{eq:SMdL/dPl}
\begin{split}
\frac{d\delta E}{d P_l} &= -\gamma u_D + a(\epsilon) P_l - b P_l^3 + P_l^5\\
&=-\gamma u_D + P_l(P_l^4 - b P_l^2 + a(\epsilon)).\\
\end{split}
\end{equation}
To find the critical endpoint, we find the smallest value of $a(\epsilon)$ for which $P_l(P_l^4 - b P_l^2 + a(\epsilon))$ is
strictly increasing for any $P_l$. We have
\begin{equation} \label{eqDerivative of dL/dPl}
\begin{split}
\frac{d}{dP_l}(P_l(P_l^4 - b P_l^2 + a(\epsilon))) &= 5 P_l^4 - 3 b P_l^2 + a(\epsilon) \\
&=5(P_l^2 - \frac{3b}{10})^2 + a(\epsilon) - \frac{9 b^2}{20} \geq 0. \\
\end{split}
\end{equation}
The smallest $a(\epsilon)$ for which the above is always positive gives the critical $a(\epsilon) = 9b^2 / 20$, at which the above derivative vanishes if we set $P_l$ to $\sqrt{3b/10}$. The value of $\gamma u_D$ at the critical endpoint is then calculated by requiring the derivative of $\delta E$ to vanish at this $a(\epsilon)$ and $P_l$:
\begin{equation} \label{eqCritialuD}
\begin{split}
\gamma u_D &= P_l(P_l^4 - b P_l^2 + a(\epsilon))\\
&=\sqrt{\frac{3}{10} b} (\frac{9}{100}b^2 -b \frac{3b}{10} + \frac{9}{20}b^2) = \frac{3\sqrt{30}}{125} b^{\frac{5}{2}}.\\
\end{split}
\end{equation}

\begin{figure}[h]
\centering
\includegraphics[width=\linewidth]{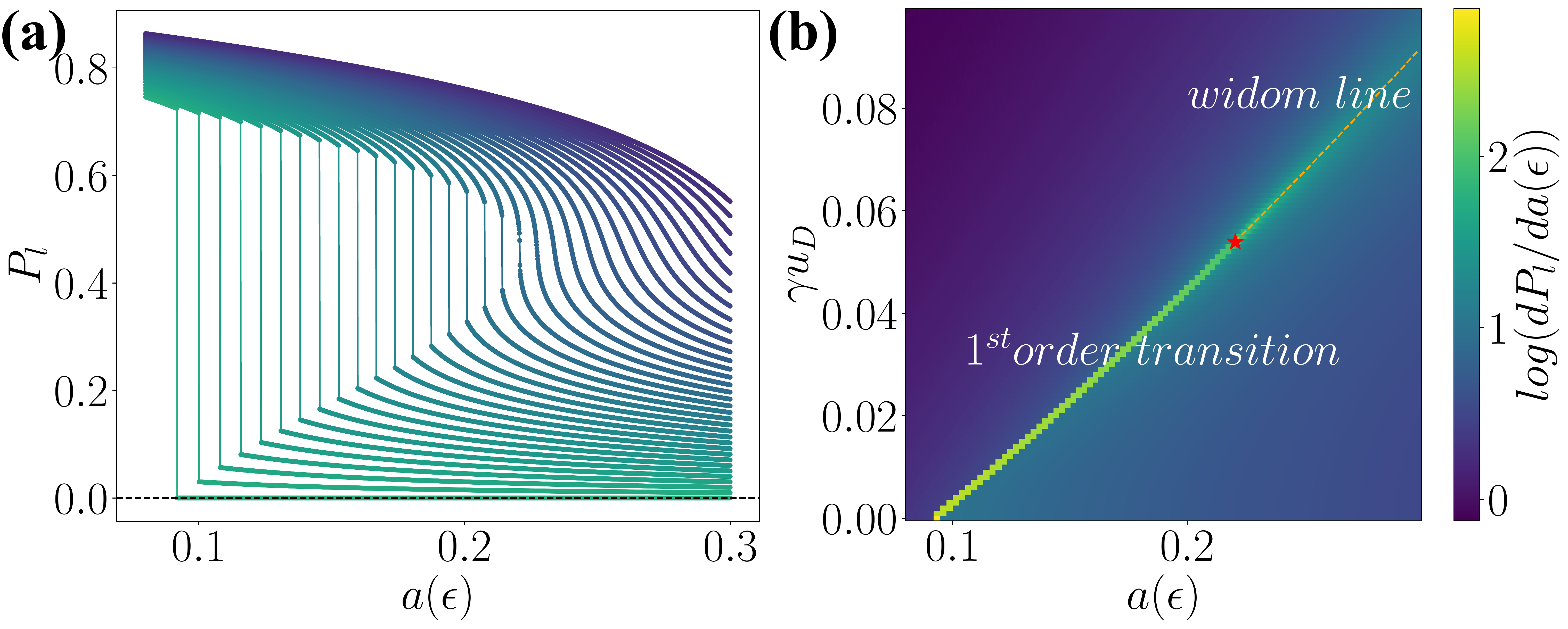}
\caption{\label{fig:smLandau} Order parameter $P_l$ with varying $a(\epsilon)$ and $\gamma u_D$ in Landau theory. These two plots capture the results obtained by finite magnetic field Hartree Fock in the main text. The parameter $b$ is fixed at $b = 0.7$ to make $P_l \in [0, 1)$. Analytic expression of critical end point(red star) is $(a(\epsilon), \gamma u_D) = (9b^2 / 20, 3 \sqrt{30} b^{5/2} / 125)$. At $\gamma u_D = 0$, the first order transition happens at $a(\epsilon) = 3 b^2 / 16$.}
\end{figure}

\section{Hybrid Wannier states at $B= 0$}

Hybrid Wannier states at $B = 0$ are eigenstates of operator $\hat{O} = \hat{P} e^{-i \frac{1}{N_1} \bG_1 \cdot \br} \hat{P}$\cite{kang2020non, wang2023revisiting}. Here $\hat{P} = \sum_{n, \bk} \ket{\psi_{n, \bk}} \bra{\psi_{n, \bk}}$ is the projector onto Bloch bands of interest.
 Since $[\hat{O}, \hat{T}(\ba_2)] = 0$, $k_2$ remains as a good quantum number, $\hat{T}(\ba_2) \ket{\myw(k_2)}=e^{-i2\pi k_2}\ket{\myw(k_2)}$. Suppose $\ket{\myw(k_2)}$ is an eigenstate of $\hat{O}$, then $\hat{T}(\ba_1) \ket{\myw(k_2)}$ is also an eigenstate, with eigenvalue shifted:
\begin{equation} \label{eq:myHWSta1}
\begin{gathered}
\hat{O} \ket{\myw(k_2)} = e^{-i \frac{1}{N_1} \langle\bG_1 \cdot \br\rangle} \ket{\myw(k_2)}\\
\hat{O} \hat{T}(\ba_1) \ket{\myw(k_2)} = \hat{T}(\ba_1) \hat{O} e^{-i \frac{1}{N_1} \bG_1 \cdot \ba_1} \ket{\myw(k_2)} = e^{-i \frac{1}{N_1} \bG_1 \cdot \ba_1} e^{-i \frac{1}{N_1} \langle\bG_1 \cdot \br\rangle} \hat{T}(\ba_1)\ket{\myw(k_2)}.
\end{gathered}
\end{equation}
Therefore we can denote all eigenstates as $\ket{\myw_{\alpha}(n_0, k_2)} = \hat{T}^{n_0}(\ba_1) \ket{\myw_{\alpha}(0, k_2)}$, where $\ket{\myw_{\alpha}(0, k_2)}$ are all eigenstates whose eigenvalues satisfy $-\pi \leq \langle \bG_1 \cdot \br \rangle < \pi = \frac{\bG_1 \cdot \ba_1}{2}$. We show the evolution of $\langle \bG_1 \cdot \br \rangle$ with varying $k_2$, for $\hat{P}_1 = \sum_{\bk} \ket{\psi_{n=1, \bk}} \bra{\psi_{n=1, \bk}}$ and $\hat{P}_2 = \sum_{\bk} \ket{\psi_{n=2, \bk}} \bra{\psi_{n = 2, \bk}}$ in Fig.~\ref{fig:myHWS}(a); the top most valance band ($\hat{P}_1$, blue) and the band below it ($\hat{P}_2$, orange) at $\theta = 3.89^{\circ}$ in $\bK$ valley of tMoTe$_2$. The winding correponds to their Chern numbers $+1$ and $-1$ respectively. Then we show the evolution for $\hat{P}_3 = \sum_{n = \{1, 2\}, \bk} \ket{\psi_{n, \bk}} \bra{\psi_{n, \bk}}$ in Fig.~\ref{fig:myHWS}(b). Corresponding plots for real space probability density in each layer of $\ket{\myw_{\alpha}(n_0 = 0, k_2 = 0)}$ is shown. All hybrid Wannier states $\ket{\myw_{\alpha}(n_0, k_2)}$ are localized along $\ba_1$ but extended along $\ba_2$. In addition, hybrid Wannier state $\ket{\myw_{\alpha}(n_0, k_2)}$ can be expressed in terms of Bloch states $\psi_{n, \bk}$ via a unitary transformation $U^{hWS}$ at each $\bk$ point
\begin{equation} \label{eq:myHWSinBloch}
\ket{\myw_{\alpha}(n_0, k_2)} = \frac{1}{\sqrt{N_1}} \sum_{n = {1, 2}} \sum_{k_1} e^{-i2 \pi k_1 n_0} U_{n, \alpha}^{hWS}(\bk) \ket{\psi_{n, \bk}}.
\end{equation}
\begin{figure}
\centering
\includegraphics[width=\linewidth]{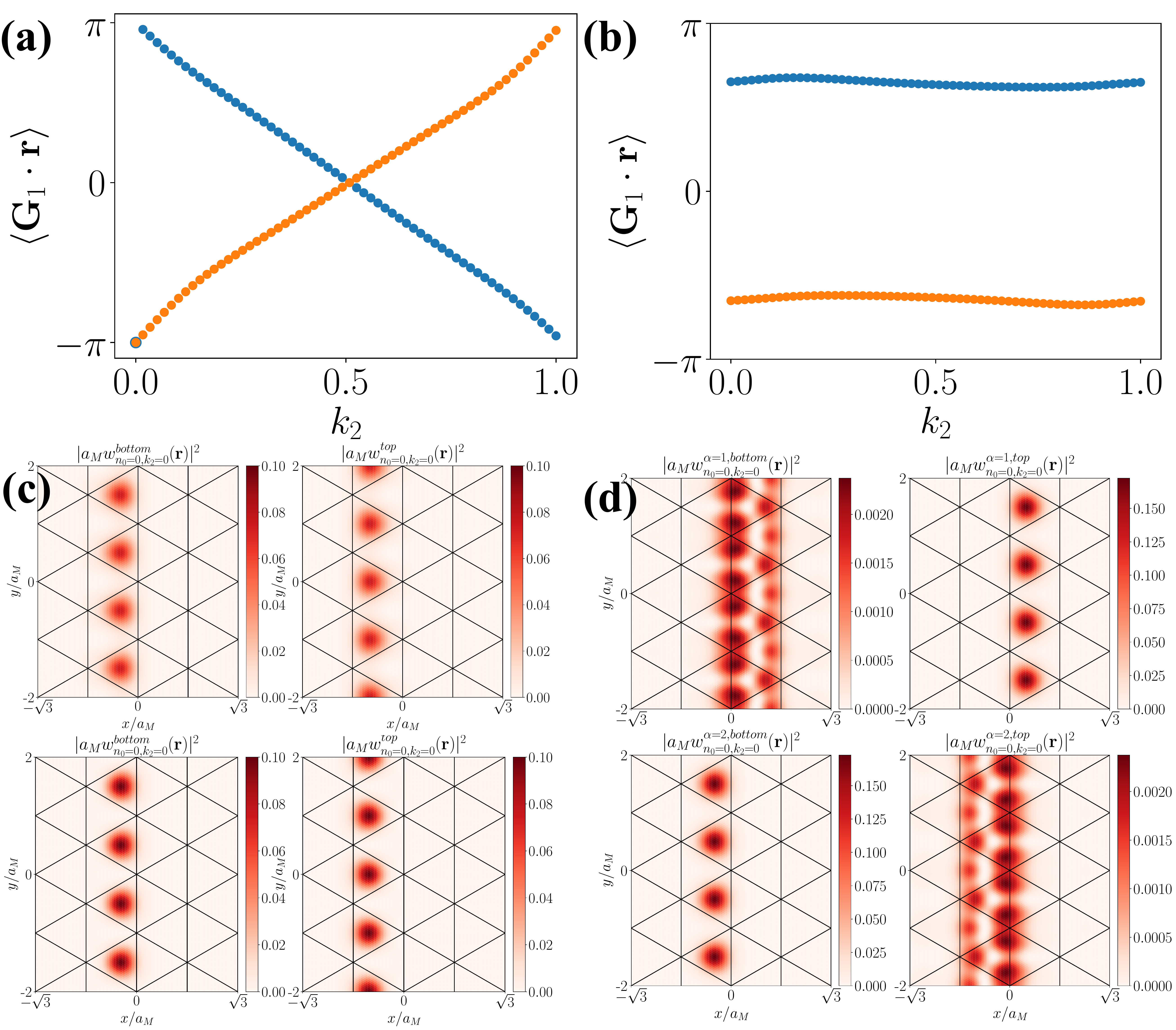}
\caption{(a, b)Average of $\langle\mathbf{\bG}_1 \cdot \mathbf{r}\rangle$ of hybrid Wannier states. (a) blue(orange) points denote the average value for hWS constructed from the first(second) band of tMoTe$_2$ in $\bK$ valley. (b)average value for two hWSs constructed from first two bands. (c) up:Probability density of $\ket{\myw(n_0 = 0, k_2 = 0)}$ from first band in $\bK$ valley. down:Probability density of $\ket{\myw(n_0 = 0, k_2 = 0)}$ from second band in $\bK$ valley. (d)Probability density of $\ket{\myw_1(n_0 = 0, k_2 = 0)}$ and $\ket{\myw_2(n_0 = 0, k_2 = 0)}$ from first two bands in $\bK$ valley.}
\label{fig:myHWS}
\end{figure}

\section{Constructing hWS basis at $B\neq0$}

In order to perform the Hartree Fock calculations at $B\neq 0$ we construct the projector onto the subspace spanned by the eigenstates of the non-interacting Hamiltonian with minimal coupling to $B$.
Specificly, we are interested in all eigenstates at $B\neq0$ that evolve from the first two valence bands in each valley at $B = 0$. The exact projector can be obtained from complete Landau level(LL) wavefunctions\cite{wang2023revisiting}. This work adapts hybrid Wannier states at $B = 0$ to approximate this projector at low magnetic field, with excellent accuracy shown in the Fig.~\ref{fig:myComparisonHF}(see Ref.~\cite{wang2023revisiting} and Ref.~\cite{wang2022narrow} for further discussion). This method allows us to perform the Hartree Fock calculations at low $B$ more efficiently than the direct Landau level expansion. In the Landau gauge $\bA = (0, Bx)$ and low $B$, hybrid Wannier states $\ket{\myw_{\alpha}(n_0=0, k_2)}$ (which are constructed at $B=0$) have an approximately same expectation value of the Hamiltonian (or any polynomial function of it) because they only extend in $\hat{y}$ and by being localized in the $\hat{x}$ near origin the region of space where the vector potential $\bA$ is large never contributes. Therefore, $\ket{\myw_{\alpha}(n_0=0, k_2)}$ must be almost completely within the $B\neq0$ Hilbert subspace evolving from the narrow bands of interest. To generate the rest of the basis at $B \neq 0$, we can employ the magnetic translation group at rational magnetic flux ratios $\phi/\phi_0=p/q$. Eigenstates of the magnetic translations $\hat{t}(\ba_1)$ and $\hat{t}(\ba_2)$ can therefore be obtained as
\begin{equation} \label{eq:myHWSwithBneq0}
\ket{W_{\alpha, r}(\bk)} = \frac{1}{\sqrt{N_1}} \sum_{s = \lfloor-\frac{N_1}{2}\rfloor}^{\lfloor \frac{N_1}{2} \rfloor} e^{i2\pi k_1 s} \hat{t}^{s}(\ba_1) \ket{\myw_{\alpha}(0, k_2 + \frac{r}{q})}.
\end{equation}
Here $\ket{\myw_{\alpha = \{1, 2\}}(0, k_2 + \frac{r}{q})}$ are constructed from $\hat{P}_3$ which projects onto the band composite with zero total Chern number. Here $k_1 \in [0, 1)$. $k_2\in [0, \frac{1}{q})$ and $r = 0, 1, ..., q - 1$ is additional index. The total number of states
at zero and finite B are the same, as must be the case because the total Chern number vanishes. We note in passing that for projectors $\hat{P}_1$ and $\hat{P}_2$ whose total Chern numbers are nonzero, we need an additional procedure to construct proper basis at $B \neq 0$ \cite{wang2023revisiting, wang2022narrow}. Under the action of $\hat{t}(\ba_1)$, $\hat{t}^q(\ba_2)$ and $\hat{t}(\ba_2)$, the basis $\ket{W_{\alpha, r}(\bk)}$ satisfies
\begin{equation} \label{eq:myWunderMangeticTranslation}
\begin{gathered}
\hat{t}(\ba_1) \ket{W_{\alpha, r}(\bk)} = e^{-i 2\pi k_1 } \ket{W_{\alpha, r}(\bk)}\\
\hat{t}^{q}(\ba_2) \ket{W_{\alpha, r}(\bk)} = e^{-i 2\pi k_2 q}\ket{W_{\alpha, r}(\bk)}\\
\hat{t}(\ba_2) \ket{W_{\alpha, r}(\bk)} = e^{-i 2\pi (k_2 + \frac{r}{q})} \ket{W_{\alpha, r}(\bk + \frac{p}{q}\bG_1)}.\\
\end{gathered}
\end{equation}
The last line follows from $\hat{t}(\ba_2) \hat{t}(\ba_1)= e^{i 2\pi \frac{\phi}{\phi_0}}\hat{t}(\ba_1) \hat{t}(\ba_2) $.

For $\frac{\phi}{\phi_0} = \frac{p}{q}$, we impose a requirement on the discrete k-space mesh at $B = 0$ so that $N_1$ is evenly divided by $2q$ and $N_2$ is evenly divided by $q$. This originates from representation of $\bq_{\phi}$ in hexagonal (triangular) lattice
\begin{equation} \label{eq:myRepresentOFqphi}
\bq_{\phi} = \frac{p}{q} \frac{2\pi}{a_M} \hat{y} = \frac{p}{q} \frac{a_{1y}}{a_M} \bG_1+ \frac{p}{q} \bG_2 = -\frac{p}{2q}\bG_1 + \frac{p}{q} \bG_2
\end{equation}
consequently, the even division is needed to represent the operator $e^{-i \bq_\phi \cdot \br}$. This is crucial to construct hybrid Wannier basis at $B \neq 0$.

Using $\ket{W_{\alpha, r}(\bk)}$, the matrix elements of the Hamiltonian can be easily evaluated\cite{wang2023revisiting}. However, it is worth mentioning that $\ket{W_{\alpha, r}(\bk)}$ are not guaranteed to be orthonormal (although they nearly are at low $B$). Therefore, a transformation is needed to generate an orthonormal basis out of them. This will mix indices $\{ \alpha, r \}$ to index $a$, with $dim(a) = dim(\alpha) dim(r)$\cite{wang2023revisiting, wang2022narrow}.
The orthonormal basis which we use is therefore
\begin{equation} \label{eq:}
\ket{V_a(\bk)} = \sum_{\alpha, r} \ket{W_{\alpha, r}(\bk)} \left(U\frac{1}{\sqrt{D}}\right)_{\alpha r, a}\!\!\!\!\!\!\!\!(\bk),
\end{equation}
where $U$ and $D$ are respectively unitary matrix and diagonal matrix of the overlap matrix $\Lambda_{\alpha, \beta, r_1, r_2} \equiv \langle W_{\beta, r_2}(\bk) |W_{\alpha, r_1}(\bk)\rangle$, namely $\Lambda = U D U^{\dagger}$.

\begin{figure}
\centering
\includegraphics[width=\linewidth]{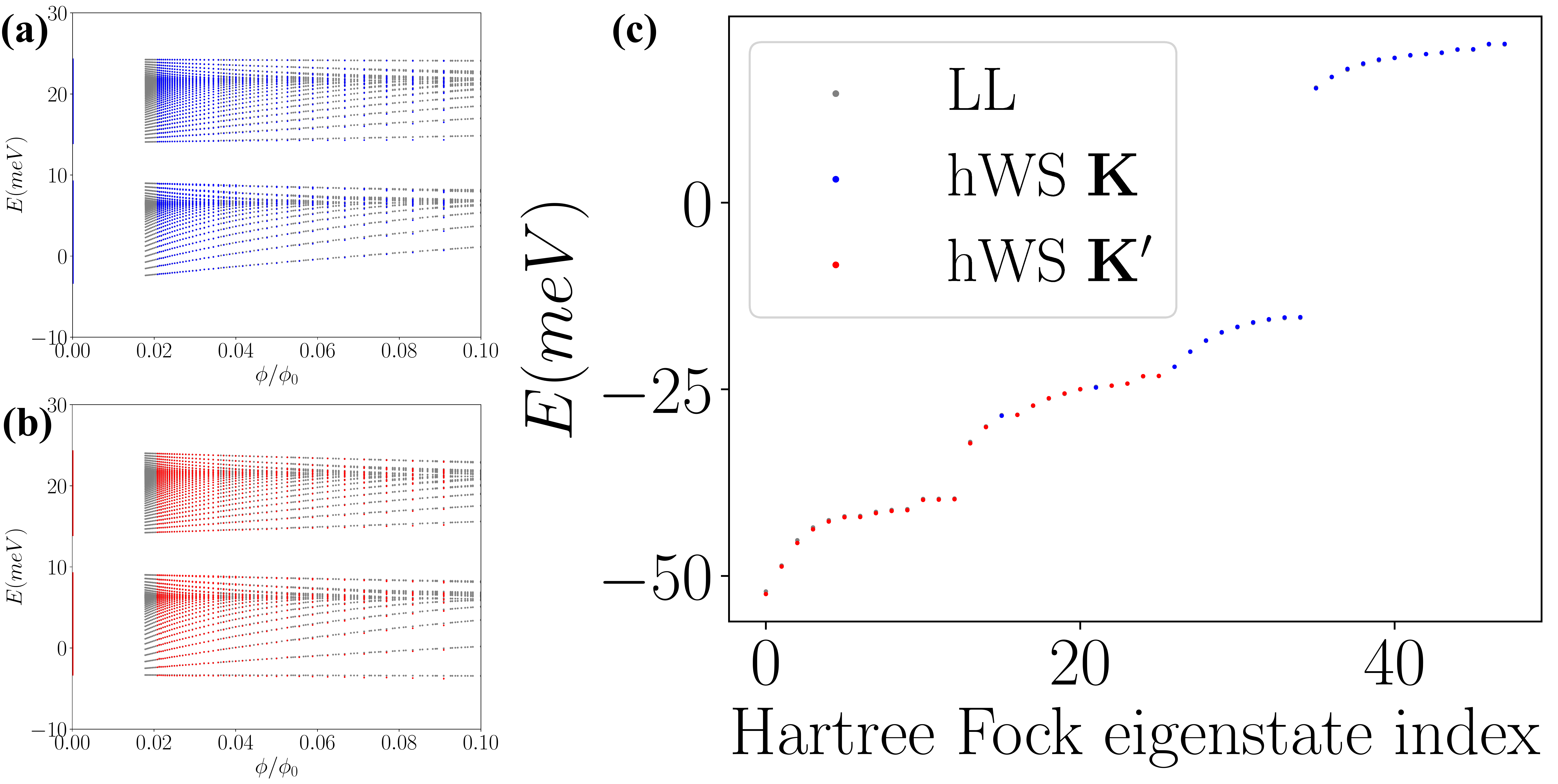}
\caption{Comparison of the Hofstadter spectra using exact LL approach and approximate hybrid Wannier state(hWS) approach. (a) Non-interacting Hamiltonian in the $\bK$ valley (b) Non-interacting Hamiltonian in the $\bK'$ valley (c) Hartree Fock mean field Hamiltonian at $\frac{\phi}{\phi_0} = \frac{1}{12}$. The good agreement demonstrates the accuracy of the hWS approach.}
\label{fig:myComparisonHF}
\end{figure}

\section{$\hat{t}(\ba_2)$ symmetry}

During our calculation, we choose points as $k_1 = 0, \frac{1}{q}, ..., \frac{q -1}{q}$ and $k_2 = 0$(Fig.~\ref{fig:myMBZschematic}). We observe that $\hat{t}(\ba_2)$ symmetry is preserved down to flux $\phi/\phi_0 = 1/18$, which we interpret to mean that ground states in TMD along (-1, -1), (-1, 0) and (-1, 1) Streda line do not break the $\hat{t}(\ba_2)$ symmetry. The similar phenomena is also observed through $B\neq 0$ Hartree-Fock calculations in twisted bilayer graphene without strain \cite{wang2023theory}. Therefore, we extend the Hartree Fock calculation down to $\phi / \phi_0 = 1/47$ by assuming that $\hat{t}(\ba_2)$ symmetry is preserved, which allows us to significantly speed up the computation at such low $B$.

\section{Extended interaction diagrams}

In extended interaction diagrams, we present, among other results, valley polarization $P_{\eta} \equiv  \frac{A_{uc}}{A}\sum_{l} \int \mathrm{d}^2 \br \langle \psi_{l \bK}^{\dagger}(\br)\psi_{l\bK}(\br) - \psi_{l\bK'}^{\dagger}(\br) \psi_{l\bK'}(\br) \rangle$. $P_{\eta} = 1$ for $\bK$ valley polarzied states, and $P_{\eta} = -1$ for $\bK'$ valley polarized states.

\begin{figure}
\centering
\includegraphics[width=\linewidth]{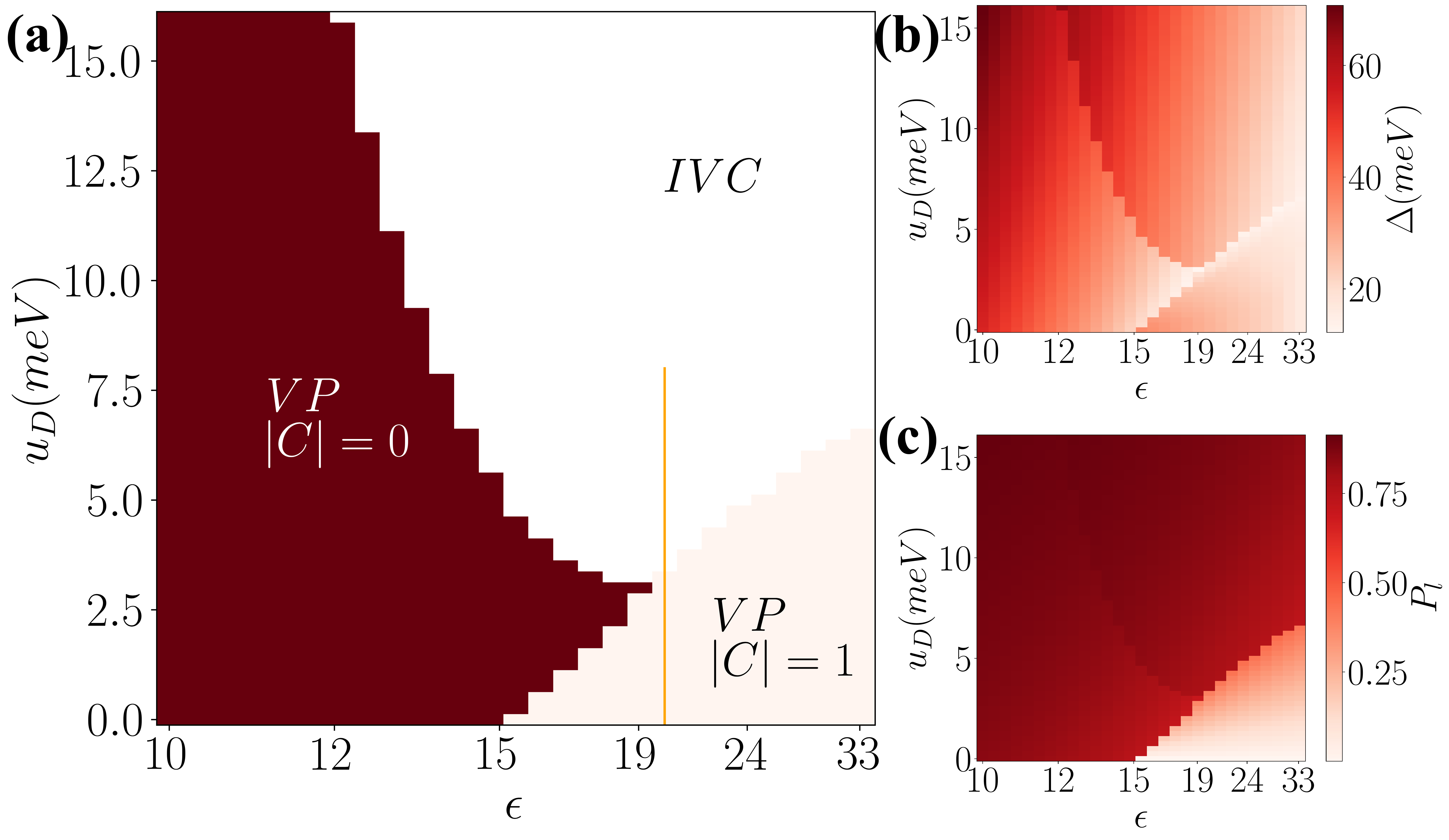}
\caption{(a)Interacting diagram at $B = 0$ at integer filling $\nu = -1$\cite{wang2023topology}. Orange line at $\epsilon = 20$ intersects the $B = 0$ topological transition from a Chern $\pm1$ insulator to a trivial insulator with a vanishing Chern number but with intervalley coherence. The behavior along such line, at the same $\epsilon$, but at $B = 8.78T$ is presented in the main text. Panels (b-c) show the charge gap and layer polarization at $B=0$, respectively. }
\label{fig:myZeroDiagram}
\end{figure}

\begin{figure}
\centering
\includegraphics[width=\linewidth]{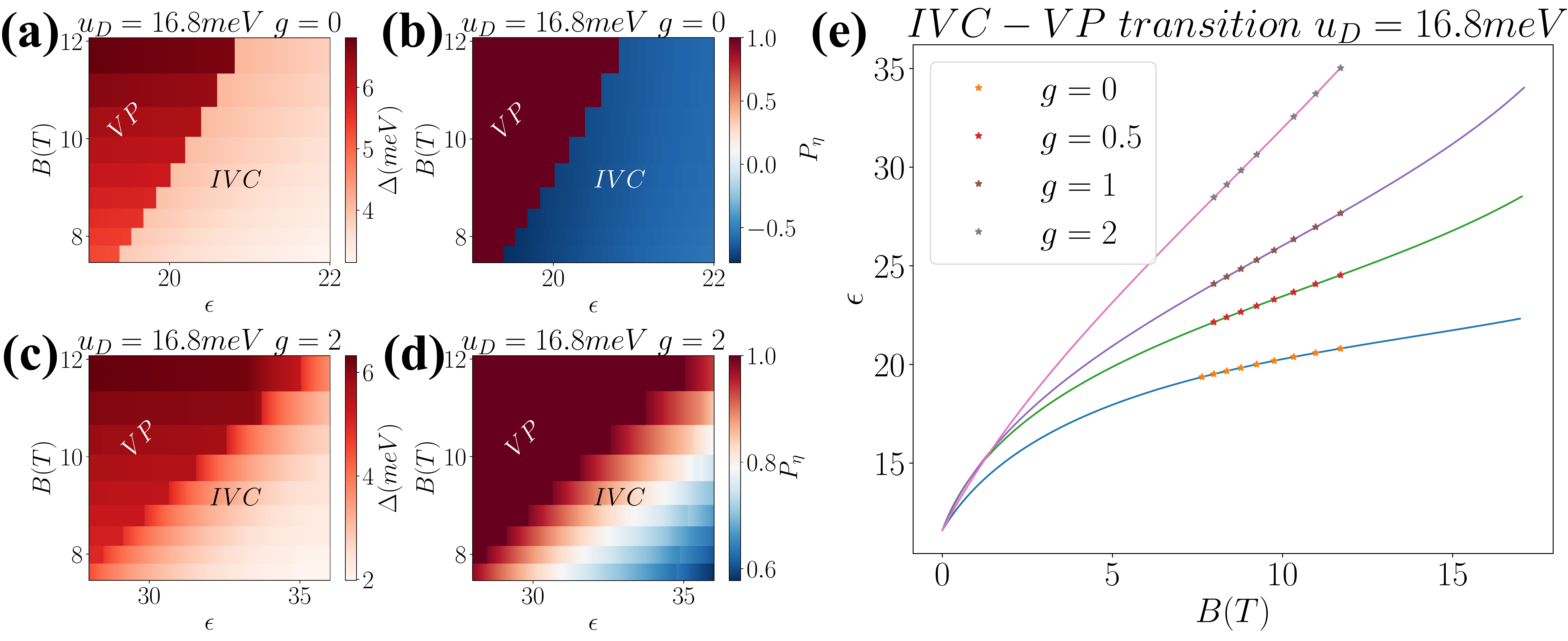}
\caption{IVC-VP transition at $u_D = 16.8meV$ with different $g$ factors. Charge gap and $P_{\eta}$ at (a,b)$g = 0$ and at (c,d)$g = 2$ are shown. (e) Extrapolated IVC-VP transition lines at various $g$. Stars denote original data points from finite magnetic field Hartree Fock calculation. }
\label{fig:MYivcVPtransition}
\end{figure}

\begin{figure}
\centering
\includegraphics[width=\linewidth]{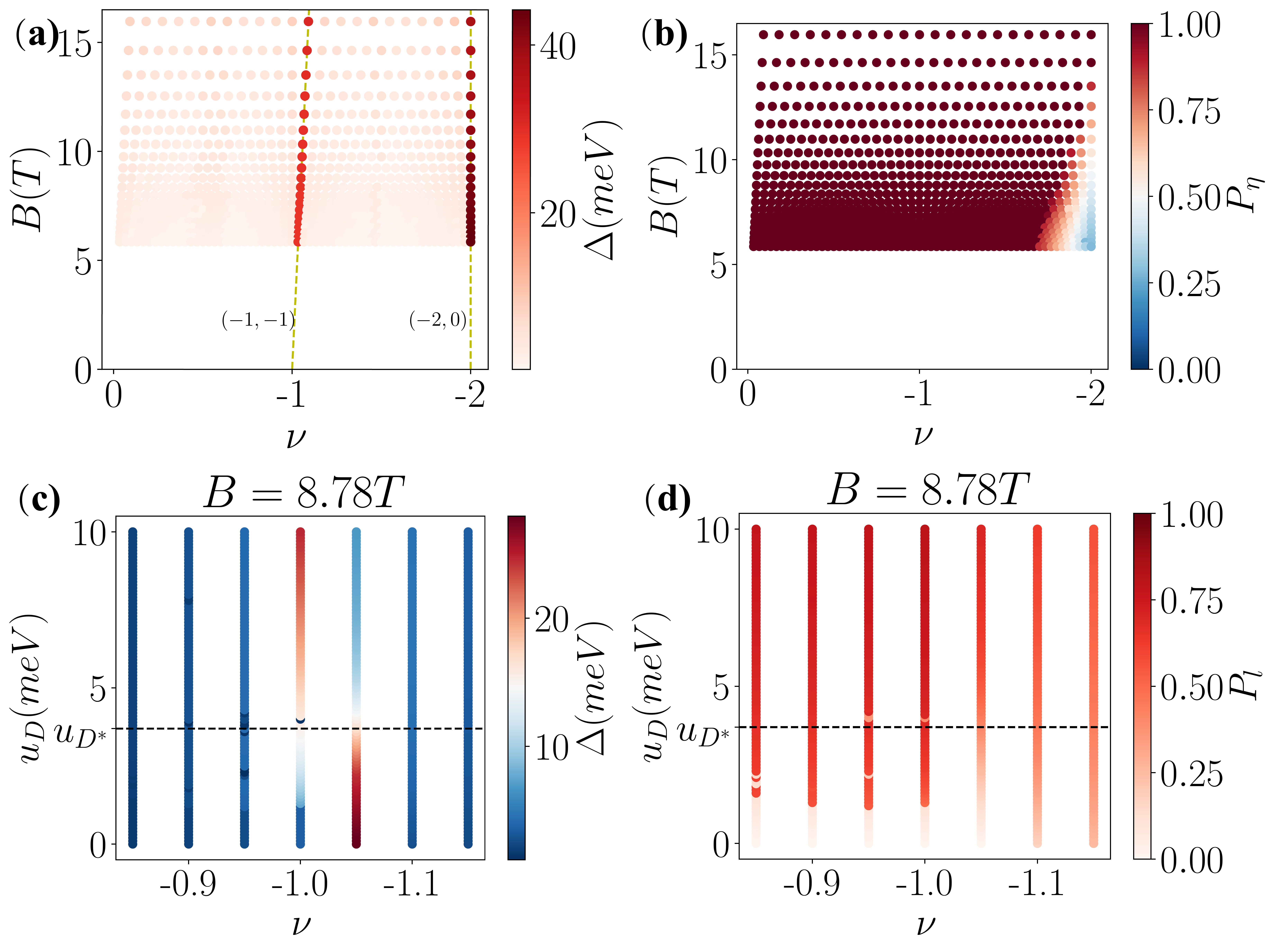}
\caption{(a-b)Charge gap and $P_{\eta}$ for various filling factors and magnetic field. (c-d)Charge gap and layer polarization $P_l$ for filling factor $\nu$ near $\nu = -1$ and various $u_D$.}
\label{fig:MYwannierANDuD}
\end{figure}

\end{widetext}
\end{document}